\documentclass[twocolumn,prl,showpacs,floatfix]{revtex4}
\usepackage{graphicx}
\usepackage{amssymb,amsmath,latexsym}
\usepackage{bm}
\usepackage{ulem}
\usepackage{bm,color}

\newcommand{\red}[1]{}

\begin{document}

\title{The coexistence of charge density wave and superconductivity in Pt-based 
	layered superconductors : SrPt$_2$As$_2$ and LaPt$_2$Si$_2$}

\author     {Sooran Kim}
\author     {Kyoo Kim}
\email[]{kyoo@postech.ac.kr}
\author     {B. I. Min} 
\email[]{bimin@postech.ac.kr}
\affiliation{Department of physics, PCTP, 
Pohang University of Science and Technology, Pohang, 790-784, Korea}



\begin{abstract}
 The intriguing coexistence of the charge density wave (CDW) and superconductivity
in SrPt$_2$As$_2$ and LaPt$_2$Si$_2$ has been investigated 
based on 
the {\it ab initio} density functional theory band structure and phonon calculations.
We have found that the local split distortions
Pt atoms in the [As-Pt-As] layers play an essential role in driving the
five-fold supercell CDW instability as well as the phonon softening instability 
in SrPt$_2$As$_2$.
By contrast, the CDW and phonon softening instabilities in LaPt$_2$Si$_2$ occur 
without split distortions of Pt atoms,
indicating that the driving mechanisms of the CDW in SrPt$_2$As$_2$ and LaPt$_2$Si$_2$
are different. 
We have found that the CDW instability for the former arises from 
the Fermi surface nesting, while, for the latter, from the saddle point scattering.
 The phonon calculations, however, suggest that 
the CDW and the superconductivity coexist in [{\it X}-Pt-{\it X}] layers 
({\it X} = As or Si) for both cases.

\end{abstract}

\pacs{71.45.Lr, 74.70.-b, 71.18.+y, 63.20.D-}
\maketitle


Low-dimensional systems often suffer from intrinsic instabilities,
revealing diverse interesting phase transitions upon cooling, such as
charge density wave (CDW), spin density wave, superconductivity (SC), and so on.
In general, those phases are detrimental to each other.
Therefore the coexistence of those phases in a system has been 
a long-standing subject of importance in the physics of low-dimensional 
systems.\cite{Gabovich01,Kiss07,Morosan06,Zhu11,Sangeetha12,Machida87} 
The Pt-based layered superconductors, SrPt$_2$As$_2$ and LaPt$_2$Si$_2$, of the present study
belong to such quasi two-dimensional (2D) systems, which exhibit the coexistence 
of the CDW and the SC at low temperature ($T$).\cite{Kudo10,Nagano13}
In fact, the Pt-based layered systems draw recent attention
because of their structural similarity to Fe-based {\it A}Fe$_2$As$_2$ (122) 
({\it A}=Ba, Ca, Sr, or Eu) superconductors, which have been intensively studied 
these days.\cite{Paglione10,Stewart11,Torikachvili08,Rotter08,Krellner08,
Kasinathan09,Takahashi08,Alireza09,Uhoya10}

\begin{figure}[b]
  \centerline{\includegraphics[width=8.5cm]{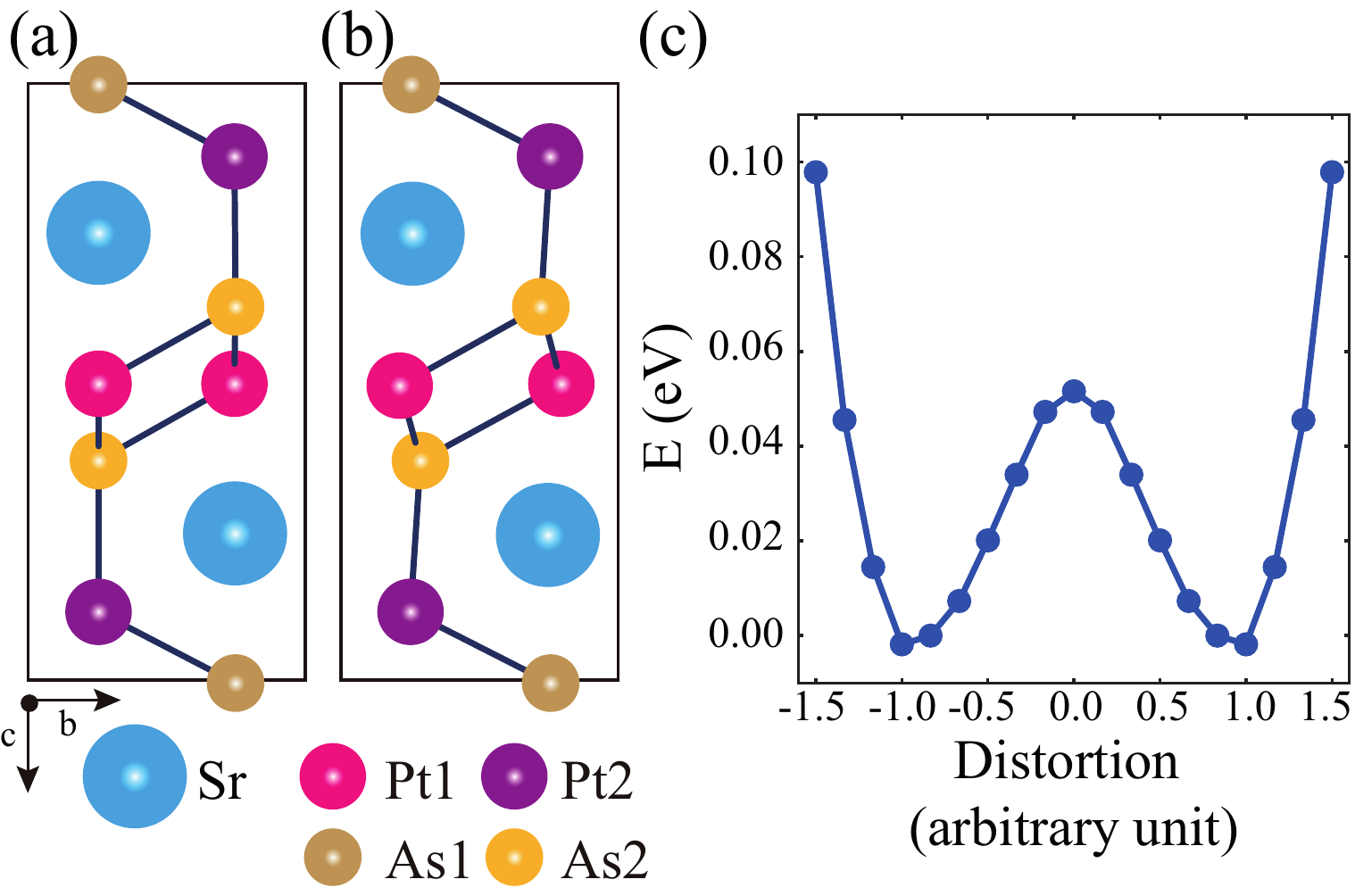}}
  \caption{
     (Color online)
    (a) Orthorhombic structure of SrPt$_2$As$_2$ with Pt1 and As2 at the  
	ideal positions without the split distortion (no-split-SPA).
    (b) Orthorhombic structure of SrPt$_2$As$_2$ with Pt1 and As2 at the
	split-off positions (split-SPA).
    (c) Total energy variation of SrPt$_2$As$_2$ with respect to the split distortion
(distortion=0.0 for no-split-SPA and distortion=1.0 for split-SPA).
}
  \label{str_ene}
\end{figure}

SrPt$_2$As$_2$ was reported recently to be
a BCS-like superconductor having two s-wave
superconducting gap feature as in MgB$_2$.\cite{Xu13}
SrPt$_2$As$_2$ undergoes a CDW transition at T$_{CDW}\simeq$ 470K,
which is accompanied by the superconducting transition at T$_c\simeq$ 5K.
Below T$_c$, the SC coexists with the CDW phase.\cite{Kudo10,Fang12}
At high $T$, SrPt$_2$As$_2$ crystallizes in the tetragonal structure 
of CaBe$_2$Ge$_2$-type ({\it P}4/{\it nmm}),
which is quite similar to ThCr$_2$Si$_2$-type structure of 
{\it A}Fe$_2$As$_2$ superconductors.\cite{Kudo10}
Differently from {\it A}Fe$_2$As$_2$, however, Pt and As in SrPt$_2$As$_2$ 
have reversed positions for every other layer, as shown in Fig.~\ref{str_ene}(a).
Namely, there are alternating [As2-Pt1-As2] and [Pt2-As1-Pt2]
layers along the $c$-direction.
The CDW modulation vector of SrPt$_2$As$_2$ was reported experimentally to be 
$q_{CDW}$=0.62 $a^*$ = (0.62, 0, 0), which yields 
the supercell structure with the modulation in the [As2-Pt1-As2] layers below T$_{CDW}$
(see Fig.~\ref{SPA_ph}).\cite{Imre07,Wang14} Even below T$_{CDW}$, 
SrPt$_2$As$_2$ has unique feature containing 
the split-off positions of Pt1 and As2, as shown in Fig.~\ref{str_ene}(b).

There have been only a few band structure calculations 
for SrPt$_2$As$_2$.\cite{Nekrasov10,Shein11}
Nekrasov {\it et al}.\cite{Nekrasov10} obtained the density of states (DOS) 
and Fermi surface (FS) of the high $T$ phase of SrPt$_2$As$_2$ 
having the tetragonal structure above T$_{CDW}$.
 They found that 5$d_{x^2-y^2}$ state of Pt1 is dominant 
at the Fermi level (E$_F$) and 
the FSs are mostly 3D-like except one cylinder-like FS.
Shein {\it et al.}\cite{Shein11} investigated the energetics of
three types of 122 system:
  CaBe$_2$Ge$_2$-type and two hypothetical ThCr$_2$Si$_2$-type structures.
They reported that CaBe$_2$Ge$_2$-type is more stable than ThCr$_2$Si$_2$-type polymorphs. 
However, none of these studies explored the electronic structures of 
the low $T$ phase of SrPt$_2$As$_2$ having 
the split-off positions of Pt and the CDW modulated structure.

Another Pt-based layered system LaPt$_2$Si$_2$ has similar physical properties 
to SrPt$_2$As$_2$.
At high $T$, LaPt$_2$Si$_2$ crystallizes in the tetragonal structure 
of CaBe$_2$Ge$_2$-type, which is similar to Fig.~\ref{str_ene}(a)
(Sr and As are replaced by La and Si).
Upon cooling, it  undergoes the CDW transition at 112 K 
with the CDW vector 
of $q_{CDW}=(n/3, 0, 0)$ ($n$=1 or 2).\cite{Nagano13}
Nagano {\it et al}.{\cite{Nagano13} suggested 
a CDW-induced supercell at low $T$, 
which corresponds to the tripling of the original unit cell.
Below T $\sim$ 2 K, the SC emerges 
in coexistence with the CDW state.\cite{Nagano13}
It was also reported that LaPt$_2$Si$_2$ is more stable in the
CaBe$_2$Ge$_2$-type structure than in the ThCr$_2$Si$_2$-type structure.\cite{Hase13}
The FSs of CaBe$_2$Ge$_2$-type structure
are mostly 2D-like, while the FSs of ThCr$_2$Si$_2$-type structure are 3D-like.
However, the unique feature of the split-off positions of Pt1 and As2 
in SrPt$_2$As$_2$ has not been 
observed in LaPt$_2$Si$_2$.

Despite existing studies on SrPt$_2$As$_2$ and LaPt$_2$Si$_2$, 
there are important remaining issues.
There has been no theoretical explanation
on the mechanisms of the observed CDW instabilities in SrPt$_2$As$_2$ and LaPt$_2$Si$_2$.
Kudo {\it et al.}\cite{Kudo10} once stated that the CDW transition
of SrPt$_2$As$_2$ seemed to originate from the FS nesting, 
but they did not specify which band is responsible for the FS nesting.
Above all, it has not been clarified
whether the CDW instabilities in SrPt$_2$As$_2$ 
and LaPt$_2$Si$_2$ have the same mechanisms or not. 
Also, there have been no phonon studies on SrPt$_2$As$_2$ and LaPt$_2$Si$_2$, 
which can provide direct clue to the CDW structural transitions.
On the basis of phonon studies, one can also investigate
SC properties in these CDW systems.

In this letter, to address the above questions, 
we have investigated the CDW and SC properties of SrPt$_2$As$_2$ and
LaPt$_2$Si$_2$, 
using the first-principles density-functional theory (DFT) band structure 
and phonon calculations.  In SrPt$_2$As$_2$, the split 
distortions of Pt1 in [As2-Pt1-As2] layers are found to
play an essential role in driving the CDW instability.
This feature in SrPt$_2$As$_2$ is 
distinct from that in LaPt$_2$Si$_2$
that does not need the split distortions to drive the CDW instability.
The phonon studies revealed that the SC emerges mainly in the CDW layer
for both SrPt$_2$As$_2$ and LaPt$_2$Si$_2$.


For the total energy band structure calculations,
the full-potential linearized augmented plane wave band method implemented in
Wien2k package was employed.\cite{W2k}
The generalized-gradient approximation (GGA) was used for the 
the exchange correlation and the spin-orbit coupling (SOC) was included.
For structural optimizations and phonon calculations,
the pseudo-potential band method implemented in VASP\cite{VASP} 
and phonopy\cite{phonopy} were used, respectively.
The supercell approach with finite displacements based on the
Hellmann-Feynman theorem\cite{Parlinski97} was used to obtain the force constants.
The pseudo-potential band method implemented in Quantum Espresso was also used 
to determine the electron-phonon (e-ph) coupling constant $\lambda_p$
and superconducting parameters.\cite{QE}

To examine the mechanism of CDW instability in SrPt$_2$As$_2$,
we focused on the role of split distortions of Pt, and so considered two structures.
  The first one is the orthorhombic structure in Fig. \ref{str_ene}(a)
without the split distortions of Pt1 and As2 (we call it ``no-split-SPA" hereafter).
 The no-split-SPA structure is close to the tetragonal CaBe$_2$Ge$_2$-type 
structure above T$_{CDW}$.  
The second one is another orthorhombic structure in Fig. \ref{str_ene}(b),
which has the split-off positions of Pt1 and As2 (hereafter ``split-SPA").
Split-SPA has the structure that is close to that of SrPt$_2$As$_2$ below T$_{CDW}$,
but does not contain the modulation by {\bf q$_{CDW}$}=0.62$a^*$.
The split-SPA structure is obtained by making the antiferro-like distortions 
	of Pt1 and As2 
and then performing the atomic relaxation.
 The initial structure data for SrPt$_2$As$_2$ 
 before the structural relaxation was adopted from
 Imre {\it et al}..\cite{Imre07} 
The initial lattice constant and atomic positions of LaPt$_2$Si$_2$ 
were adopted from Shelton {\it et al}.,\cite{Shelton84} and
Nekrasov {\it et al}.,\cite{Nekrasov10} respectively.
  The relaxed structural parameters of SrPt$_2$As$_2$ and
  LaPt$_2$Si$_2$ are summarized in the supplement.\cite{Supp}


\begin{figure}[b]
  \centerline{\includegraphics[width=8.5cm]{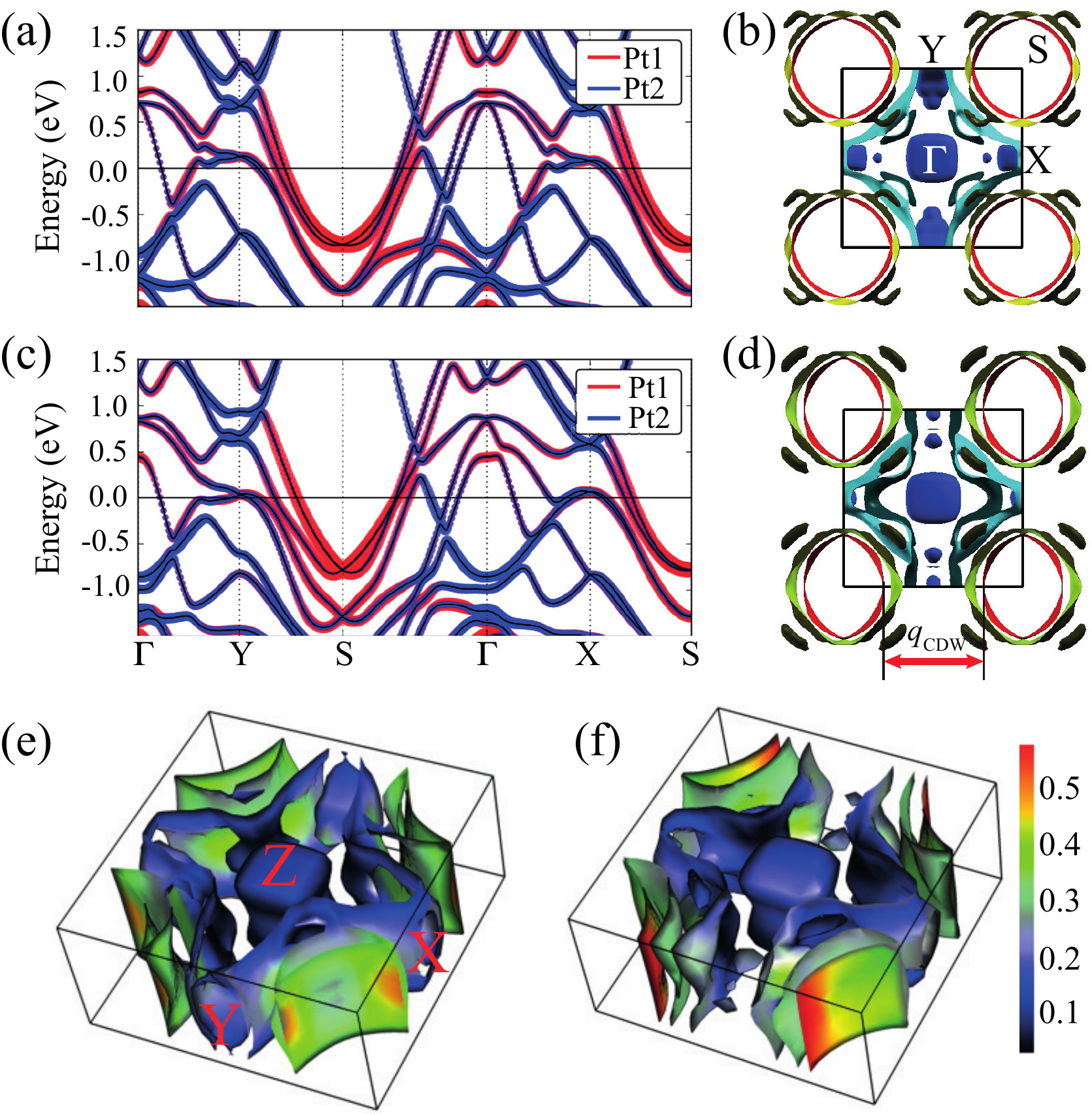}}
  \caption{
     (Color online)
(a) Band structure of no-split-SPA.
     Pt1 and Pt2 band characters are shown with fat bands.
(b) FS of no-split-SPA in the $a^*b^*$ plane.
(c) Band structure of split-SPA.
(d) FS of split-SPA in the $a^*b^*$ plane. 
	The red FSs centered at S become flatter, 
	which provides the nesting vector q$_{CDW}$.
(e) FS of no-split-SPA in the full Brillouin zone (BZ).
(f) FS of split-SPA in the full BZ.
      Color bar for (e),(f) represents the Pt1 contribution to the FS (maximum value=1).
For split-SPA, the flat-red region is seen to be enhanced at the S-centered FSs.
  }
  \label{band_fs}
\end{figure}

First, we checked the energetics of SrPt$_2$As$_2$ with respect to
the split distortion.
  The total energy variation from no-split-SPA to split-SPA 
is shown in Fig. \ref{str_ene}(c).
The negative distortion means the opposite split directions of Pt1 and As2.
  We obtained the double-well shaped energy profile, 
which indicates that 
the split distortions of Pt1 and As2 indeed lower the total energy.
The energy difference between no-split-SPA and split-SPA 
is $\Delta E$ $\simeq$ 50 meV/u.c..

 Figure \ref{band_fs} shows band structures and FSs of no-split-SPA and split-SPA.
  As shown in Fig. \ref{band_fs}(a) and (c), the main character of 
the dispersive bands around S is attributed to Pt1 band 
in both no-split-SPA and split-SPA.
  But the significant difference between the no-split-SPA and 
split-SPA is revealed in the FSs. 
  Figure \ref{band_fs}(b) and (e) for no-split-SPA show mainly the 3D-like FSs 
 except one cylinder-like FS centered at S
(red-colored FS in Fig. \ref{band_fs}(b)), 
as is consistent with existing calculations.\cite{Nekrasov10,Shein11}
  Interestingly, the circular-cylindrical FS for no-split-SPA
is changed into the 
ellipsoidal-cylindrical FS for split-SPA, 
	 as shown in Fig. \ref{band_fs}(d),
	and so the parallel portion of the FS is increased.
Pt1-projected FSs of no-split-SPA and split-SPA are presented in 
Fig. \ref{band_fs}(e) and (f), respectively.  
For no-split-SPA, the FS has almost 4-fold rotational symmetry,
  and the Pt1 projection is distributed rather uniformly over the cylindrical FS.
On the other hand, 
for split-SPA, the 4-fold rotational symmetry is completely broken because 
of the ellipsoidal-cylindrical FSs at S. 
It is amusing to find in Fig. \ref{band_fs}(d) that the nesting vector connecting 
the flat parts of ellipsoidal FSs is in good agreement with 
the experimental CDW modulation vector of $q_{CDW}$= (0.62, 0, 0)
suggested by Imre {\it et al}..\cite{Imre07}
  This result demonstrates that the split distortions of Pt1 and As2 
in [As2-Pt1-As2] layer of split-SPA are essential to drive the CDW transition.
Also notable in Fig. \ref{band_fs}(f) is that the Pt1 character is
dominant at longer parts of the ellipsoidal FSs, which clearly indicates that 
the Pt1 band is responsible for the CDW instability in split-SPA.

\begin{figure}[t]
  \centerline{\includegraphics[width=8 cm]{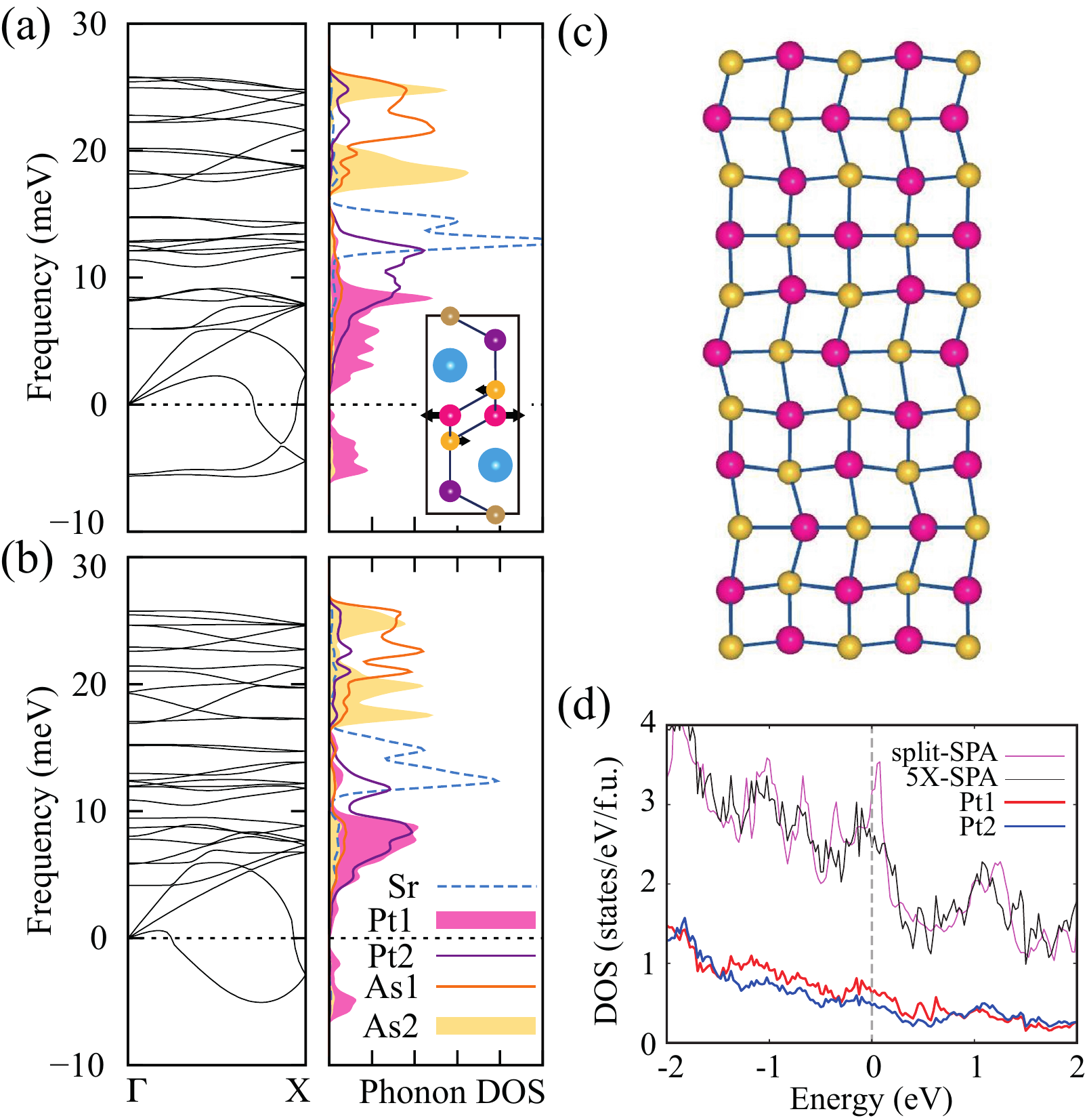}}
  \caption{
     (Color online)
(a) Phonon dispersion and partial phonon DOS for no-split-SPA. 
(b) The same for split-SPA.
The negative frequency here implies the imaginary phonon frequency, 
	indicating the structural instability.
In the inset of (a), the softened phonon mode at $\Gamma$ is depicted.
(c) The modulated structure (5X-SPA) of [As2-Pt1-As2] layer in the {\it ab} plane due to 
	the softened phonon mode at {\bf q}=(0.6, 0, 0) of (b).
(d) Total electron DOSs are compared between the split-SPA and 5X-SPA cases. 
	Partial electron DOSs of Pt1 and Pt2 of 5X-SPA are also plotted.
  }
  \label{SPA_ph}
\end{figure}

 In order to check the structural instability more directly,
we performed phonon dispersion calculations 
for both no-split-SPA and split-SPA.
As shown in Fig. \ref{SPA_ph}(a) and (b),
the phonon softening instabilities occur in both cases, 
indicating the structural instabilities.
This feature is consistent with experiment in that 
the ground state of SrPt$_{2}$As$_{2}$ has the CDW modulated structure.\cite{Imre07}
The softened phonon modes arise mainly from Pt1,
as shown in the partial phonon DOSs, suggesting that the CDW transition occurs
in the Pt1 layers.
Inset of Fig. \ref{SPA_ph}(a) shows the normal mode of softened phonon
at $\Gamma$ for no-split-SPA, 
which induces the split distortions 
of Pt1 and As2. This phonon mode induces the structural transition 
from no-split-SPA to split-SPA, 
which is consistent with structural energetics
in Fig. \ref{str_ene}(c).

  Figure \ref{SPA_ph}(b) shows that the $\Gamma$ point softening disappears for split-SPA,
 and the phonon softening instability becomes dominant near $q_{CDW}$=(0.62, 0, 0).
 Indeed, the relaxed structure modulated by the softened phonon mode at {\bf q}=(0.6, 0, 0)
in Fig. \ref{SPA_ph}(c) is very close to the experimentally 
suggested structure.\cite{Imre07}
 We will refer to this relaxed structure as 5X-SPA,
	 as it is five-fold supercell structure due to {\bf q}=(0.6, 0, 0).
 Figure \ref{SPA_ph}(d) provides the total DOSs of split-SPA and 5X-SPA,
which shows that 5X-SPA is still metallic, 
reflecting that the CDW nesting is imperfect.
This is one reason why the CDW and the SC could coexist 
in SrPt$_2$As$_2$.\cite{Gabovich01}
The DOS at E$_F$ is lower for 5X-SPA than for split-SPA.
 Notable feature is that, even after the CDW transition,
the contribution to the DOS at E$_F$ comes more from Pt1 band.
  The ratio of Pt1 and Pt2 DOSs at E$_F$ is $\sim$1.3.
 This suggests that the [As2-Pt1-As2] layer would be more susceptible to
the emergence of the SC.
 We will discuss this point more below.

\begin{figure}[t]
  \centerline{\includegraphics[width=8.5cm]{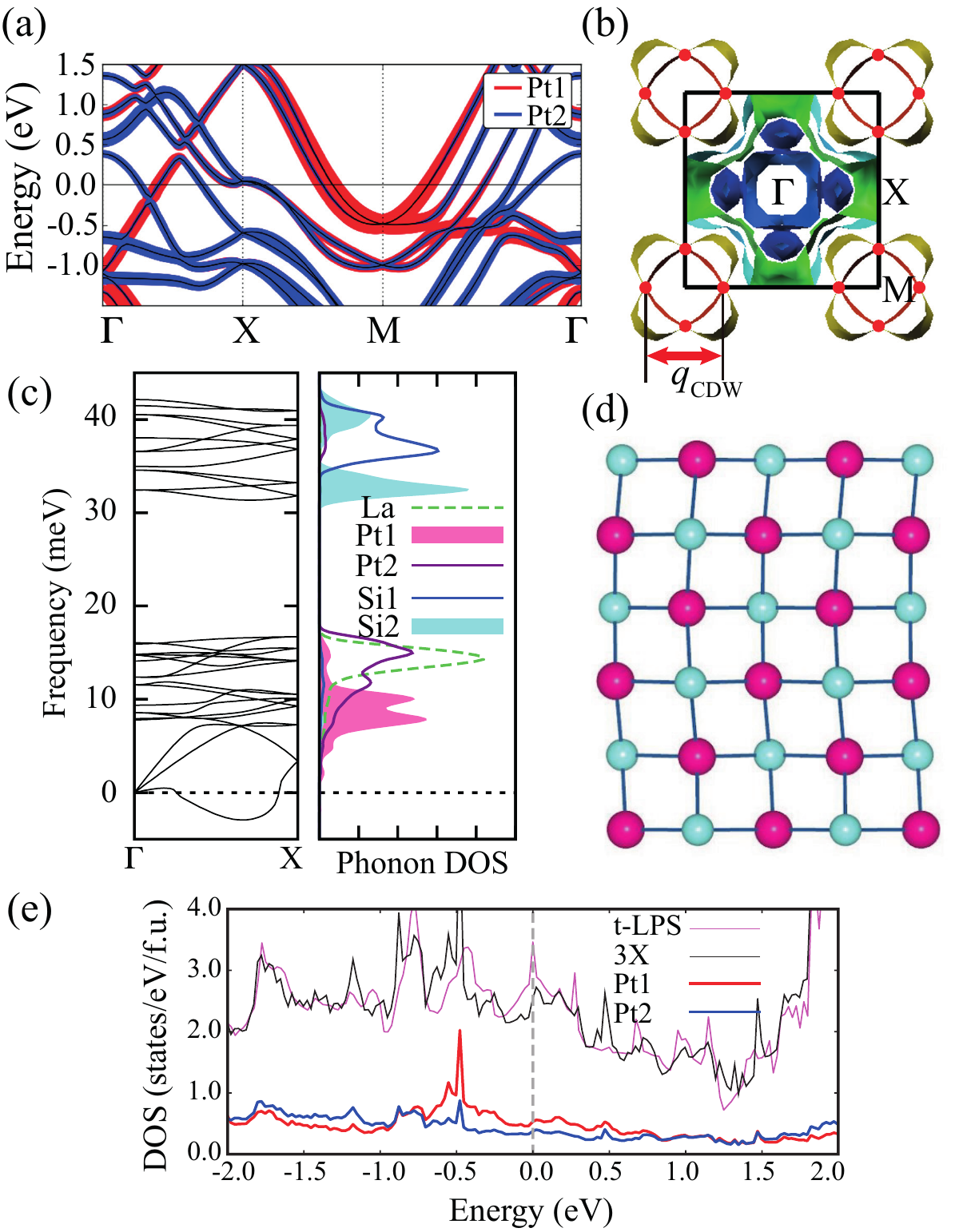}}
  \caption{
     (Color online)
    (a) Band structure of tetragonal LaPt$_2$Si$_2$ (t-LPS).
    Pt1 and Pt2 band characters are shown with fat band.
    (b) FS of t-LPS. Red points represent the saddle points.
    The scattering vector connecting the saddle points
 is close to the observed q$_{CDW}$.
    (c) Phonon dispersion and partial phonon DOS of t-LPS.
    (d) The modified structure (3X-LPS) of [Si2-Pt1-Si2] layer in the {\it ab} plane
     due to the softened phonon mode at {\bf q}=(1/3, 0, 0). 
    (e) Total electron DOSs are compared between t-LPS and 3X-LPS.
    Partial electron DOSs of Pt1 and Pt2 of 3X-LPS are also plotted.
}
  \label{LPS_ph}
\end{figure}

For comparison, we performed the band structure and phonon calculations for
tetragonal LaPt$_2$Si$_2$ (t-LPS).
Figure \ref{LPS_ph}(a) shows the band structure of t-LPS.
The Pt1 band produces the electron pocket FSs around M, 
which possess the saddle points, colored in red, in Fig. \ref{LPS_ph}(b).
The scattering vector connecting the saddle points is indeed close 
to the observed $q_{CDW}$=(1/3, 0, 0).\cite{Nagano13}
In fact, the saddle point scattering 
produces a peak in the charge susceptibility,
which induces the structural instability.\cite{Rice75,Xu87}

 The phonon dispersion of t-LPS in Fig.~\ref{LPS_ph}(c) 
contains the softened phonon mode,
which is consistent with the CDW structural transition.
 The phonon softening occurs mainly from the Pt1, as in SrPt$_2$As$_2$.
But, in contrast to the case in SrPt$_2$As$_2$, the phonon softening at $\Gamma$  
causing the split distortions of Pt1 does not occur.
It is compatible with the nonexistence of the split distortion in LaPt$_2$Si$_2$.

The difference between LaPt$_2$Si$_2$ and SrPt$_2$As$_2$ 
in the split distortions is expected to arise from 
the volume difference between SrPt$_2$As$_2$ 
(100.08 \AA$^3$/f.u.)
and LaPt$_2$Si$_2$ (89.62 \AA$^3$/f.u.).
We found that the split distortions in SrPt$_2$As$_2$ disappear  
with decreasing the volume.
In contrast, the split distortions appear in LaPt$_2$Si$_2$
with increasing the volume.
These features suggest that stability of split distortions 
is strongly dependent on the volume of the system.
Detailed volume-dependent behaviors are provided in the supplement.\cite{Supp}
We note another Pt-based superconductor, BaPt$_2$Sb$_2$,
which also contains the split-off Pt positions in [Sb-Pt-Sb] layer.\cite{Imai14} 
The volume of BaPt$_2$Sb$_2$ is 118.47 \AA$^3$/f.u., which is 
larger than those of SrPt$_2$As$_2$ and LaPt$_2$Si$_2$.

 The phonon softening instability at $q$ = (1/3, 0, 0) is consistent 
	with the observed CDW vector $q_{CDW}$ = (1/3, 0, 0),
which produces the three-fold supercell structure, as shown in Fig. \ref{LPS_ph}(d).
    We will refer to this structure as 3X-LPS.
The modulations occur mainly in Pt1 layer of [Si2-Pt1-Si2], as in SrPt$_2$As$_2$.
 This result is contrary to the speculation of Nagano {\it et al.},\cite{Nagano13}
 who claimed that [Pt2-Si1-Pt2] layer would be the CDW layer.

 Figure \ref{LPS_ph}(e) presents the DOSs of t-LPS and 3X-LPS.
It is seen that 3X-LPS is still metallic even with the CDW distortion.
The DOS at E$_F$ is lower for 3X-LPS than for t-LPS,
which is consistent with the stabilized 3X-LPS and also with the
paramagnetic susceptibility measurement.\cite{Nagano13}
 The ratio of Pt1 and Pt2 DOS at E$_F$ for 3X-LPS is $\sim$1.4.
The higher DOS at E$_F$ for Pt1 suggests that Pt1 layer is more
susceptible to the SC transition.

To identify the main superconducting layer, 
we performed the calculation 
of e-ph coupling constant, $\lambda_p$, for 3X-LPS.\cite{Supp}
The largest $\lambda_{q}(\nu)$ is obtained at $q=\Gamma$ near $\nu=$3.9 meV
(see supplement).\cite{Supp}
The normal mode at this frequency is mainly composed of
displacements of Pt1 
in [Si2-Pt1-Si2] layer, which suggests that
the main contribution to the SC comes from Pt1 layers.
Noteworthy is that the Pt1 layer is the CDW-modulated layer,
which implies that the SC and the CDW coexist in the same layer.
The arrangement of Pt in [Si2-Pt1-Si2] layer is more 2D-like than in [Pt2-Si1-Pt2] layer. 
It is thus expected that the 2D nature and the CDW modulation of [Si2-Pt1-Si2] layer 
facilitate the emergence of the SC more effectively.
The situation in SrPt$_2$As$_2$ is thought to be the same as in LaPt$_2$Si$_2$
in view of the structural and electronic similarities below T$_{CDW}$.

Finally, we have evaluated the superconducting parameters for 3X-LPS 
	using the Eliashberg e-ph coupling theory
 and the Allen-Dynes formula for the critical temperature $T_c$.\cite{Eliashberg60,Allen75}
The results for the Eliashberg function, $\alpha^2F(\omega)$, 
and the electron-phonon coupling constant, $\lambda_p(\omega)$, of 3X-LPS
are provided in the supplement.\cite{Supp}
We have obtained T$_c$=3.5 K for $\mu^*$ = 0.13 
($\mu^*$: the effective Coulomb repulsion parameter),
which is in good agreement with the observed $T_c$ of $\sim$ 2 K.\cite{Nagano13,Shelton84}


In conclusion, we have found that the CDW mechanisms are quite different 
between SrPt$_2$As$_2$ and LaPt$_2$Si$_2$.
The CDW transition in SrPt$_2$As$_2$ arises from the 
FS  nesting in the presence of the split distortions of Pt1 atoms,
while that in LaPt$_2$Si$_2$ from the saddle points scattering
in the absence of the split distortions.
These features are corroborated by the phonon softening instabilities
at the observed CDW modulation vector of q$_{CDW}$.
In both cases, however, Pt1 band plays an essential role in the CDW and 
superconducting transitions, 
implying that the CDW and the SC coexist in the Pt1 layers.

This work was supported by the NRF (Grant No. 2011-0025237), POSTECH BSRI grant,
 and KISTI (Project No. KSC-2013-C3-065).


\newpage

\setcounter{table}{0}
\setcounter{figure}{0}
\onecolumngrid

\begin{center}
{\bf \Large
{\it Supplemental Material:}\\
The coexistence of charge density wave and superconductivity in Pt-based 
        layered superconductors : SrPt$_2$As$_2$ and LaPt$_2$Si$_2$
}
\end{center}

\author     {Sooran Kim}
\author     {Kyoo Kim}
\author     {B. I. Min} 
\affiliation{Department of physics, PCTP,
Pohang University of Science and Technology, Pohang, 790-784, Korea}

\twocolumngrid

\section{Structural parameters}
\begin{table}[h]
\centering
\caption{Structural parameters of 
SrPt$_2$As$_2$ without split distortions of Pt1 and As2 atoms 
(space group \it{Pmmn}).
}
\begin{ruledtabular}
\begin{tabular}{cccc}
\multicolumn{4}{c}{a=4.482 \AA, b= 4.525 \AA, c=9.869 \AA}\\
     site     &   {\it x} & {\it y}  & {\it z} \\ \hline
     Sr      & 0.75000  & 0.25000  & 0.24690 \\
     Pt1     & 0.75000  & 0.75000  & 0.49890 \\
     Pt2     & 0.75000  & 0.25000  & 0.88170 \\
     As1    & 0.75000  & 0.75000  & 0.99970 \\
     As2    & 0.75000  & 0.25000  & 0.62630 \\
 \end{tabular}
 \end{ruledtabular}
\label{SPA1}
\end{table}
\begin{table}[h]
\centering
\caption{
Structural parameters of SrPt$_2$As$_2$ with split distortions 
of Pt1 and As2 atoms (space group {\it P}-1).
}
\begin{ruledtabular}
\begin{tabular}{cccc}
\multicolumn{4}{c}{a=4.482 \AA, b= 4.525 \AA, c=9.869, $\alpha$=$\beta$=$\gamma$=90$^{\circ}$}\\
     site     &   {\it x} & {\it y}  & {\it z} \\ \hline
     Sr      & 0.75000  & 0.253367  & 0.248931 \\
     Pt1     & 0.75000  & 0.795670  & 0.497525 \\
     Pt2     & 0.75000  & 0.247472  & 0.879795 \\
     As1    & 0.75000  & 0.749726  & 0.998933 \\
     As2    & 0.75000  & 0.283194  & 0.627811 \\
 \end{tabular}
 \end{ruledtabular}
\label{SPA2}
\end{table}

\begin{table}[h]
\centering
\caption{Structural parameters of LaPt$_2$Si$_2$. Space group {\it P}4/{\it nmm}}
\begin{ruledtabular}
\begin{tabular}{cccc}
\multicolumn{4}{c}{a=4.31559 (4.277) \AA, c=9.87452 (9.798)
\footnote{The values in parentheses are the initial lattice constants 
from the experiment.\cite{Shelton84}} \AA} \\
     site     &   {\it x} & {\it y}  & {\it z} 
\footnote{The values in parentheses are the initial atomic positions 
from idealized tetragonal SrPt$_2$As$_2$
before the relaxation.\cite{Nekrasov10}} \\ \hline
     La    & 0.25000  & 0.25000  & 0.743933 (0.74690) \\
     Pt1     & 0.25000  & 0.75000  & 0.000000 (0.00000) \\
     Pt2     & 0.25000  & 0.25000  & 0.379697 (0.38170) \\
     Si1    & 0.25000  & 0.75000  & 0.500000 (0.50000) \\
     Si2    & 0.25000  & 0.25000  & 0.130249 (0.12630) \\
 \end{tabular}
 \end{ruledtabular}
\label{LPS}
\end{table}
Table~\ref{SPA1} and Table~\ref{SPA2} provide
lattice parameters of SrPt$_2$As$_2$, 
which were used in the calculations.
The atomic positions are shifted by (0.5, 0, 0.5) 
from those of Imre {\it et al}..\cite{Imre07} 
The lattice parameters of LaPt$_2$Si$_2$ before and after the relaxation 
are provided in Table \ref{LPS}.

\section{Volume effect on the split distortions}

We have investigated the volume effect on the split distortions
of Pt1 and As2 (or Si2) atoms.
 We first determined the optimized atomic positions at each volume 
by performing the structural relaxation calculations.
 Figure \ref{Vol_eff} shows the split distortion sizes of Pt1 and As2 (or Si2)
as a function of volume. 
It is shown that the split distortions become stable 
with increasing the volume in both SrPt$_2$As$_2$ and LaPt$_2$Si$_2$.
It is demonstrated that the difference between SrPt$_2$As$_2$ and LaPt$_2$Si$_2$
in the split distortions originates from their different volumes.

\begin{figure}[h]
  \centerline{\includegraphics[width=8cm]{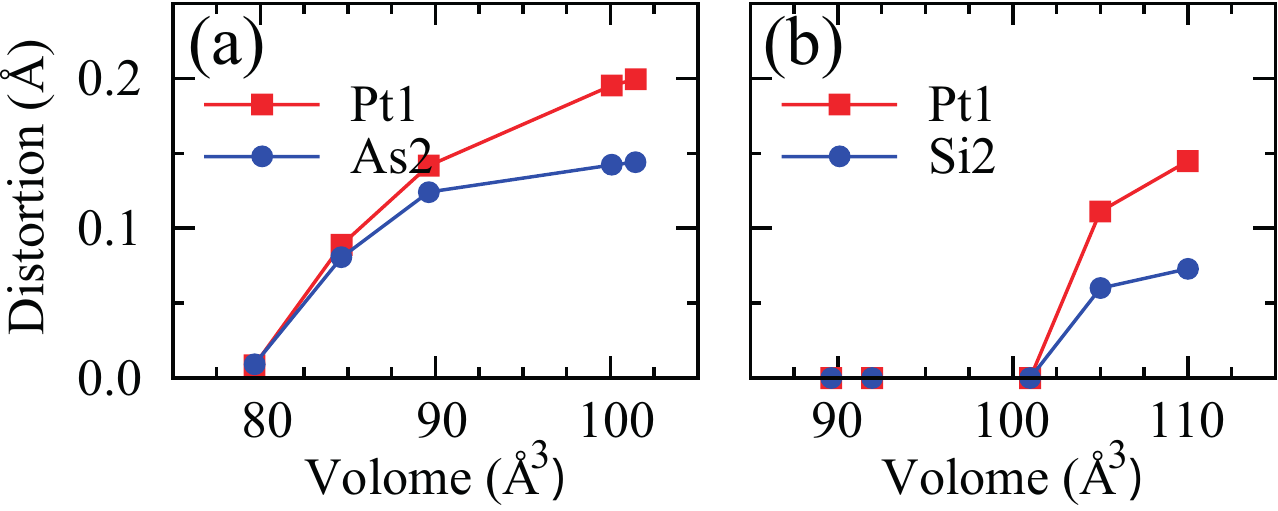}}
  \caption{
     (Color online)
    The variation of split distortion with the volume. (a) SrPt$_2$As$_2$ 
    (b) LaPt$_2$Si$_2$. The experimental volumes are 100.08 \AA$^3$/f.u.
    for SrPt$_2$As$_2$ and 89.62 \AA$^3$/f.u. for LaPt$_2$Si$_2$.
    }
  \label{Vol_eff}
\end{figure}

\section{Effect of the spin-orbit coupling on the electronic structure of 
CDW distorted LaPt$_2$Si$_2$}
\begin{figure}[h]
  \centerline{\includegraphics[width=7cm]{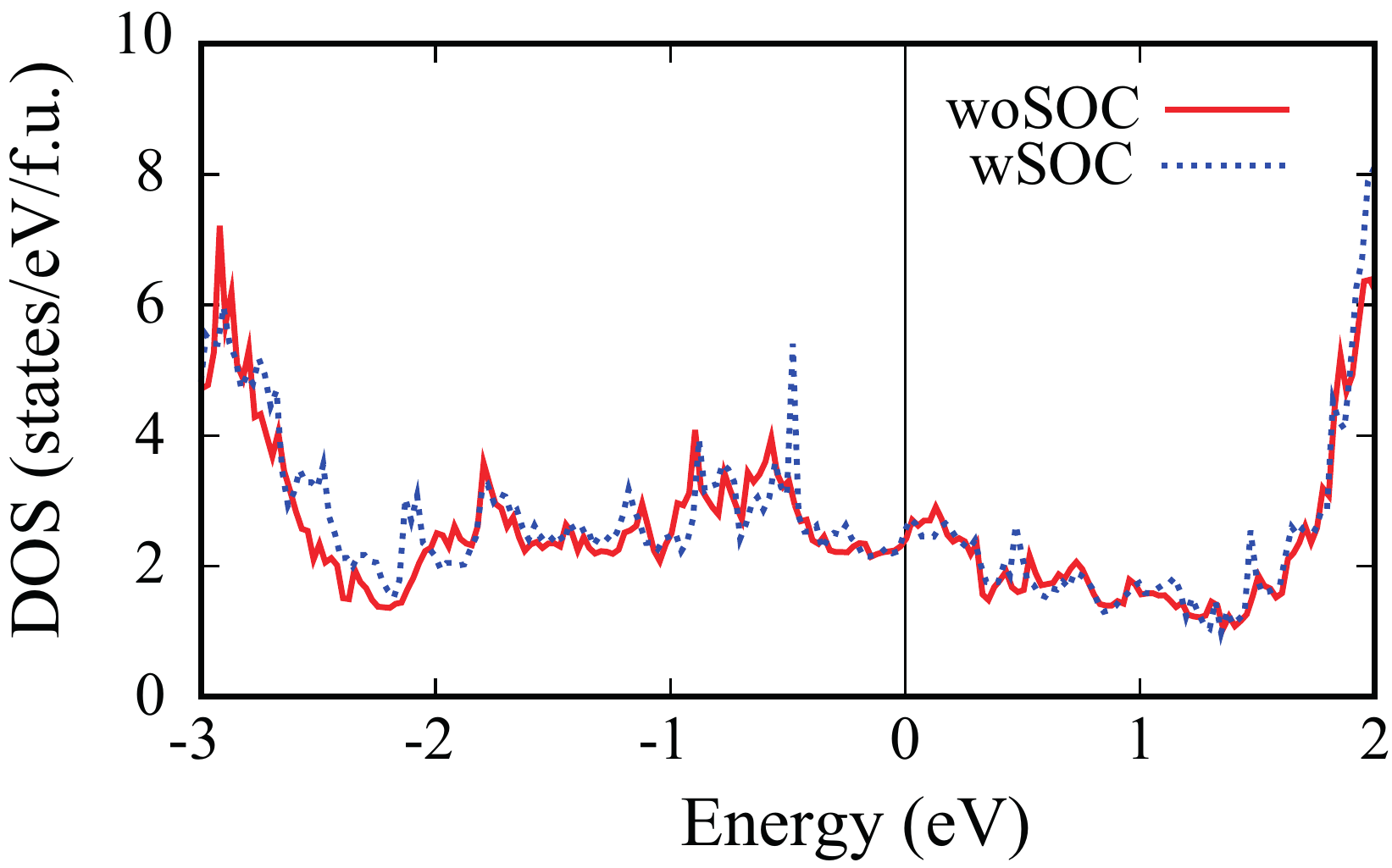}}
  \caption{
     (Color online)
    The total DOS of CDW distorted 3X-LPS.
    The red and blue dotted lines represent the total DOSs 
	with the SOC (wSOC) and without the SOC (woSOC), respectively.
}
  \label{LPS_dos}
\end{figure}
We have checked the spin-orbit coupling (SOC) effect on
the total density of states (DOS) of LaPt$_2$Si$_2$ 
after the CDW transition, which is denoted by 3X-LPS in the main text.
The total DOSs obtained with the SOC and without the SOC do not have significant 
difference, as shown in Fig. \ref{LPS_dos}.
Therefore, we carried out the electron-phonon coupling constant
calculation without the SOC.

\section{Eliashberg function, $\lambda_p(\omega)$, 
and other superconducting parameters of LaPt$_2$Si$_2$}

We have evaluated the superconducting parameters for 3X-LPS 
after the charge density wave (CDW) transition,
using the Eliashberg e-ph coupling theory
 and the Allen-Dynes formula for the critical temperature 
$T_c$,\cite{Eliashberg60,Allen75}
 \begin{equation}
 T_c = \frac{\omega_{log}}{1.20}
        \exp\left[\frac{-1.04(1+\lambda_p)}{\lambda_p(1-0.62\mu^*)-\mu^*}\right],
 \end{equation}
 where $\omega_{log}=\exp\left[\frac{2}{\lambda_p}\int\frac{d\omega}{\omega} 
 \alpha^2F(\omega)\log\omega\right]$, $\alpha^2F(\omega)$ 
 is the Eliashberg function, 
and $\mu^*$ is the effective Coulomb repulsion parameter.

Figure \ref{LPS_a2F} shows the Eliashberg function, $\alpha^2F(\omega)$,
and the electron-phonon coupling constant, $\lambda_p(\omega)$,
of 3X-LPS.
The peak of $\alpha^2F(\omega)$ and abrupt change of $\lambda_p(\omega)$ appear
at around $\sim$3.9 meV. 
Indeed this phonon frequency yields the largest contribution to $\lambda_{q\nu}$.
As provided in Table \ref{SC},
 the estimated T$_c$=3.5 K for $\mu^*$ = 0.13, is quite consistent with the
observed $T_c$ of $\sim$ 2 K.

\begin{figure}[h]
  \centerline{\includegraphics[width=8cm]{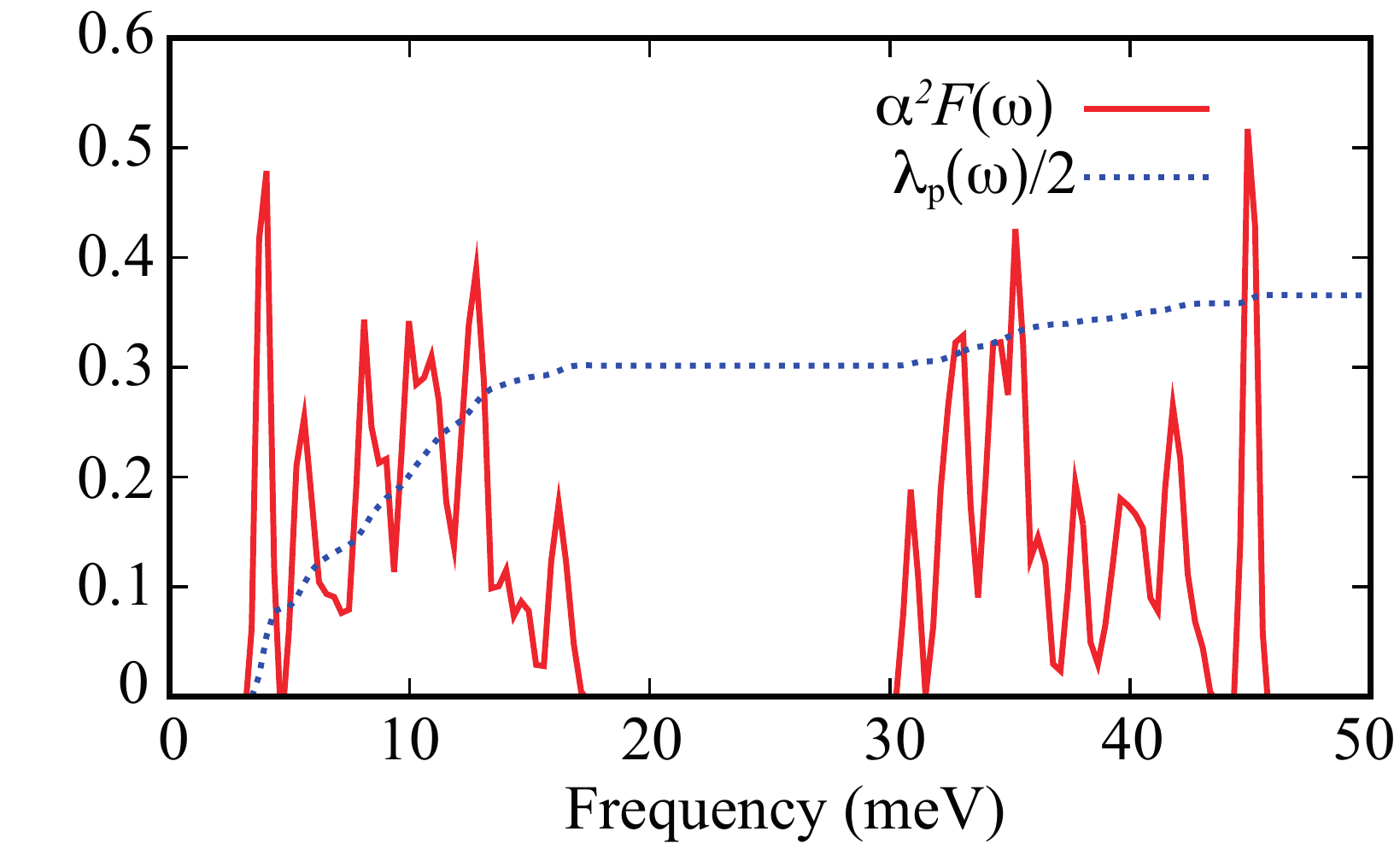}}
  \caption{
     (Color online)
     The Eliashberg function and $\lambda_p(\omega)$ of 3X-LPS.
    }
  \label{LPS_a2F}
\end{figure}
\begin{table}[h]
\centering
\caption{Superconducting parameters of LaPt$_2$Si$_2$. $N(E_F)$, $\omega_{log}$, 
and $\Theta_D$ are the DOS at $E_F$, the logarithmic average phonon frequency, 
and the Debye temperature, respectively. 
$T_c$'s are obtained for two effective Coulomb repulsion parameters 
$\mu^*$ =0.1 and 0.13.
}
\begin{ruledtabular}
\begin{tabular}{ccccc}
    $N(E_F)$  & $\omega_{log} (K)$ &   $\Theta_D$ (K) & $\lambda_p$  & $T_c$ (K) \\ 
    (states/eV/f.u.)  &     &    & & $\mu^*$ =0.1, 0.13  \\ \hline
    2.3    &  117.34 & 141.78 & 0.73  & 4.5, 3.5\\
   \end{tabular}
 \end{ruledtabular}
\label{SC}
\end{table}



\begin{thebibliography}{99}
  \bibitem{Gabovich01} A. M. Gabovich, A. I. Voitenko, J. F. Annett, and M. Ausloos,
	Supercond. Sci. Technol. {\bf 14}, R1 (2001).
   \bibitem{Kiss07} T. Kiss, T. Yokoya, A. Chainani, S. Shin, T. Hanaguri, M. Nohara, 
   and H. Takagi, Nat. Phys. {\bf 3}, 720 (2007).
   \bibitem{Morosan06}
   E. Morosan, H. W. Zandbergen, B. S. Dennis, J. W. G. Bos, Y. Onose, T. Klimczuk, 
   A. P. Ramirez, N. P. Ong, and R. J. Cava, Nat. Phys. {\bf 2}, 544 (2006).
   \bibitem{Zhu11}
    X. Zhu, H. Lei, and C. Petrovic, Phys. Rev. Lett. {\bf 106}, 246404 (2011).
   \bibitem{Machida87}
   K. Machida and M. Kato, Phys. Rev. B 36, 854 (1987).
  \bibitem{Sangeetha12}
   N. S. Sangeetha, A. Thamizhavel, C. V. Tomy, S. Basu, A. M. Awasthi, 
  S. Ramakrishnan, and D. Pal, Phys. Rev. B {\bf 86}, 024524 (2012).
   \bibitem{Kudo10}
       K. Kudo, Y. Nishikubo, and M. Nohara,
       J. Phys. Soc. Jpn. {\bf 79}, 123710 (2010)
    \bibitem{Nagano13}
      Y. Nagano, N. Araoka, A. Mitsuda, H. Yayama, H. Wada, M. Ichihara, M. Isobe, and Y. Ueda,
      J. Phys. Soc. Jpn. {\bf 82}, 064715 (2013).
   \bibitem{Paglione10}
      J. Paglione and R. L. Greene, Nat. Phys. {\bf 6}, 645 (2010).
   \bibitem{Stewart11}
      G. R. Stewart, Rev. Mod. Phys. {\bf 83}, 1589 (2011).
   \bibitem{Torikachvili08}
      M. S. Torikachvili, S. L. Bud'ko, N. Ni, and P. C. Canfield, 
	Phys. Rev. Lett. {\bf 101}, 057006 (2008).
   \bibitem{Rotter08}
      M. Rotter, M. Tegel, and D. Johrendt, Phys. Rev. Lett. {\bf 101}, 107006 (2008).
   \bibitem{Krellner08}
      C. Krellner, N. Caroca-Canales,  A. Jesche, H. Rosner, A. Ormeci, and C. Geibel,
      	Phys. Rev. B {\bf 78}, 100504(R) (2008).
   \bibitem{Kasinathan09}
      D. Kasinathan, A. Ormeci, K. Koch, U. Burkhardt, W. Schnelle, A. Leithe-Jasper,
      and H. Rosner, New J. Phys. {\bf 11}, 025023 (2009).
   \bibitem{Takahashi08}
      H. Takahashi, H. Okada, K. Igawa, K. Arii, Y. Kamihara, S. Matsuishi, M. Hirano,
      H. Hosono, K. Matsubayashi, and Y. Uwatoko, J. Phys. Soc. Jap. {\bf 77}, Suppl. C 78 (2008).
   \bibitem{Alireza09}
      P. L. Alireza, Y. T. C. Ko, J. Gillett, C. M. Petrone, J. M. Cole, G. G. Lonzarich, and S. E. Sebastian,
      J. Phys. Condens. Matter {\bf 21}, 012208 (2009).
   \bibitem{Uhoya10}
      W. Uhoya, G. Tsoi, Y. K. Vohra, M. A. McGuire, A. S. Sefat, B. C. Sales, D. Mandrus, and S. T. Weir,
      J. Phys. Condens. Matter {\bf 22}, 292202 (2010).
   \bibitem{Xu13}
       X. Xu, B. Chen, W. H. Jiao, B. Chen, C. Q. Niu, Y. K. Li,
       J. H. Yang, A. F. Bangura, Q. L. Ye, C. Cao, J. H. Dai,
       G. Cao, and N. E. Hussey,
       Phys. Rev. B {\bf 87}, 224507 (2013).
  \bibitem{Fang12}
       A. F. Fang, T. Dong, H. P. Wang, Z. G. Chen, B. Cheng,
       Y. G. Shi, P. Zheng, G. Xu, L. Wang, J. Q. Li, and N. L. Wang,
       Phys. Rev. B {\bf 85}, 184520 (2012).
   \bibitem{Imre07}
       A. Imre, A. Hellmann, G. Wenski, J. Graf, D. Johrendt, and A. Mewis,
       Z. Anorg. Allg. Chem. {\bf 633}, 2037 (2007).
   \bibitem{Wang14}
	The additional CDW modulation vector, $q$=0.23 $a^*$,
	was recently reported by
    L. Wang, Z. Wang, H.-L. Shi, Z. Chen, F.-K. Chiang, H.-F. Tian, 
    H.-X. Yang, A.-F. Fang, N.-L. Wang, and J.-Q. Li, 
	Chinese Phys. B {\bf 23}, 086103 (2014).
   \bibitem{Nekrasov10}
       I. A. Nekrasov, and M. V. Sadovskii,
       JETP letters, {\bf 92}, 751 (2010).
   \bibitem{Shein11}
       I. R. Shein, and A. L. Ivanovskii,
       Phys. Rev. B {\bf 83}, 104501 (2011).
   \bibitem{Hase13}
       I. Hase, T. Yanagisawa,
       Physica C {\bf 484}, 59 (2013).
   \bibitem{W2k}
       P. Blaha, K. Schwarz, G. K. H. Madsen, D. Kvasnicka, and
       J. Luitz, WIEN2K, An Augmented Plane Wave Plus Local Orbitals
       Program for Calculating Crystal Properties (Vienna University
       of Technology, Austria, 2001).       
   \bibitem{VASP}
       G. Kresse and J. Furthm{\"u}ller,
       Phys. Rev. B {\bf 54}, 11169 (1996); Comput. Mater. Sci. 6, 15 (1996).
   \bibitem{phonopy}
       A. Togo, F. Oba, and I. Tanaka,
       Phys. Rev. B {\bf 78}, 134106 (2008).
   \bibitem{Parlinski97}
     K. Parlinski, Z. Q. Li, and Y. Kawazoe,
     Phys. Rev. Lett. {\bf 78}, 4063 (1997).
  \bibitem{QE}{P. Giannozzi, S. Baroni, N. Bonini, M. Calandra, R. Car, C. Cavazzoni, 
  D. Ceresoli, G. L. Chiarotti, M. Cococcioni, I. Dabo, A. Dal Corso, S. Fabris, G. Fratesi, 
  S. de Gironcoli, R. Gebauer, U. Gerstmann, C. Gougoussis, A. Kokalj, M. Lazzeri, 
  L. Martin-Samos, N. Marzari, F. Mauri, R. Mazzarello, S. Paolini, A. Pasquarello, 
  L. Paulatto, C. Sbraccia, S. Scandolo, G. Sclauzero, A. P. Seitsonen, A. Smogunov, 
  P. Umari, R. M. Wentzcovitch, J. Phys. Condens. Matter {\bf 21}, 395502 (2009).}
   \bibitem{Shelton84}
      R. N. Shelton, H. F. Braun, and E. Musick, Solid State Commun. {\bf 52}, 797 (1984).
   \bibitem{Supp}{See {\it Supplemental Material} for detailed structure parameters,
  the volume effect on the split distortions, the effect of the SOC 
   in 3X-LPS, and the superconducting parameters of 3X-LPS.  }
  \bibitem{Rice75}{T. M. Rice and G. K. Scott, Phys. Rev. Lett. {\bf 35}, 120 (1975).}  
   \bibitem{Xu87}
      J.-H. Xu, T. J. Watson-Yang, J. Yu, and A. J. Freeman,
      Phys. Lett. A {\bf 120}, 489 (1987).
  \bibitem{Imai14} 
   M. Imai, S. Ibuka, N.Kikugawa, T. Terashima, S. Uji, T. Yajima, H. Kageyama, and I. Hase, 
   arXiv.1409.7147.
  \bibitem{Eliashberg60} {G. M. Eliashberg: Zh. Eksp. Teor. Fiz. {\bf 38}, 966 (1960).}
  \bibitem{Allen75} {P. B. Allen and R. C. Dynes: Phys. Rev. B {\bf 12}, 905 (1975).}
\end{thebibliography}

\begin{thebibliography}{99}
\bibitem{Imre07}
       A. Imre, A. Hellmann, G. Wenski, J. Graf, D. Johrendt, and A. Mewis,
       Z. Anorg. Allg. Chem. {\bf 633}, 2037 (2007).
\bibitem{Shelton84}
      R. N. Shelton, H. F. Braun, and E. Musick, Solid State Commun. {\bf 52}, 797 (1984).
\bibitem{Nekrasov10}
       I. A. Nekrasov, and M. V. Sadovskii,
       JETP letters, {\bf 92} 751 (2010).
\bibitem{Eliashberg60}{G. M. Eliashberg: Zh. Eksp. Teor. Fiz. {\bf 38}, 966 (1960).} 
\bibitem{Allen75}{P. B. Allen and R. C. Dynes: Phys. Rev. B {\bf 12}, 905 (1975).}
\end{thebibliography}
\end{document}